# SMILE: A novel way to explore solar-terrestrial interactions

G. Branduardi-Raymont[1], C. Wang[2]

[1]Mullard Space Science Laboratory, Department of Space and Climate Physics, University College London, Holmbury St Mary, Dorking, Surrey RH5 6NT, United Kingdom, e-mail: g.branduardi-raymont@ucl.ac.uk

[2]National Space Science Center, Chinese Academy of Sciences, No. 1 Nanertiao, Zhongguancun, Haidian district, Beijing 100190, China

**Abstract** This chapter describes the scientific motivations that led to the development of the SMILE (Solar wind Magnetosphere Ionosphere Link Explorer) mission. The solar wind coupling with the terrestrial magnetosphere is a key link in Sun-Earth interactions. In-situ missions can provide detailed observations of plasma and magnetic field conditions in the solar wind and the magnetosphere, but leave us still unable to fully quantify the global effects of the drivers of Sun-Earth connections, and to monitor their evolution. This information is essential to develop a comprehensive understanding of how the Sun controls the Earth's plasma environment and space weather. SMILE offers a new approach to global monitoring of geospace by imaging the magnetosheath and cusps in X-rays emitted when high charge-state solar wind ions exchange charges with exospheric neutrals. SMILE combines this with simultaneous UV imaging of the northern aurora and in-situ plasma and magnetic field measurements in the magnetosheath and solar wind from a highly elliptical northern polar orbit. In this chapter the science that SMILE will explore and the scientific preparations that will ensure the optimal exploitation of SMILE measurements are presented.

**Keywords**: Earth's magnetosphere, solar wind, charge exchange X-rays, imaging

## 1 Introduction

The solar wind coupling with the terrestrial magnetosphere is a key link in Sun-Earth interactions. Mass and energy enter geospace mainly via dayside magnetic reconnection under southward Interplanetary Magnetic Field (IMF) conditions; reconnection in the tail leads to release of energy and particle injection deep into the magnetosphere, causing geomagnetic storms and substorms. One end product of this is the visual manifestation of variable auroral emissions. In-situ missions can provide detailed observations of the plasma and magnetic field conditions in both the solar wind and the magnetosphere. However, we are still unable to fully quantify the global effects of the drivers of Sun-Earth connections, and to monitor their evolution with time. This information is essential to develop a comprehensive understanding of how the Sun controls the Earth's plasma environment and space weather. This is not just matter of scientific curiosity – it also addresses a clear and pressing practical problem. As our world becomes ever more dependent on complex technology – both in space and on the ground – society becomes more exposed to the vagaries of space weather, the conditions on the Sun and in the solar wind, magnetosphere, ionosphere and thermosphere that can influence the performance and reliability of technological systems and endanger human life and health.

SMILE (Solar wind Magnetosphere Ionosphere Link Explorer, Branduardi-Raymont et al. 2018) offers a new approach to global monitoring of geospace by imaging the magnetosheath and cusps in soft (low energy) X-rays emitted when high charge-state solar wind ions exchange charges with exospheric neutrals. SMILE is a self-standing mission coupling X-ray imaging of the magnetosheath and polar cusps (large spatial scales in the magnetosphere) with simultaneous UV imaging of global auroral distributions (mesoscale structures in the ionosphere) and in-situ solar wind/magnetosheath plasma and magnetic field measurements. SMILE will provide scientific data on the solar wind-magnetosphere interaction continuously for long, uninterrupted periods of time from a highly elliptical northern polar orbit. SMILE is a collaborative mission between the European Space Agency (ESA) and the Chinese Academy of Sciences (CAS), currently under development and due for launch at the end of 2024. This chapter describes the novel science that SMILE will deliver, while a chapter in Section III dedicated to 'The SMILE mission' presents the ongoing technical developments and scientific preparations, and the current status, of the mission and payload.

## 2 The Earth's magnetosphere

The interaction of the Earth's magnetic field with the super-Alfvénic and supersonic solar wind plasma leads to the formation of the magnetosphere (Fig. 1). The flow of the solar wind compresses the dayside of the magnetosphere and drags its nightside out into a long magnetotail. A collisionless bow shock forms upstream of the magnetosphere in the supersonic solar wind. The shocked solar wind plasma flows around the magnetosphere through the magnetosheath. A relatively sharp transition from dense, shocked, solar wind plasma to tenuous magnetospheric plasma marks the magnetopause. High latitude cusps denote locations where field lines divide to close either in the opposite hemisphere or far down the magnetotail. The radial magnetic field within the cusps allows the plasma to flow along the field lines and provides an opportunity for the solar wind to penetrate deep into the magnetosphere, all the way to the ionosphere. The position and shape of the magnetopause change constantly as the Earth's magnetosphere responds to varying solar wind dynamic pressure and interplanetary magnetic field strength and orientation.

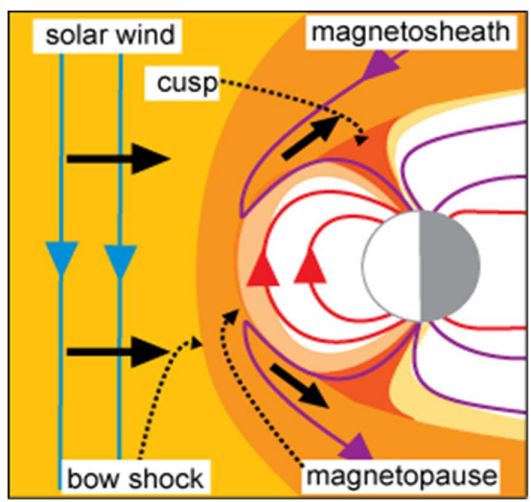

Fig. 1 A cut through the Earth's dayside magnetosphere, shown in the XZ plane. The Sun is to the left, with positive X increasing towards it and Z orthogonal to the Sun-Earth line and positive upwards. The magnetopause is the outer boundary of the magnetosphere, is compressed on the dayside and extends in a long tail on the nightside. The bow shock compresses, slows and deflects the solar wind plasma so that it can flow around the magnetopause.

Intervals of magnetic reconnection at the dayside magnetopause under southward IMF conditions lead to closed field lines becoming open and direct penetration of solar wind plasma into the magnetosphere. This removes magnetic flux from the dayside magnetosphere and adds it (and corresponding magnetic energy) to the magnetotail lobes. Here the stored energy is intermittently and explosively released, following episodes of magnetic reconnection and the closing of magnetic field lines: the process takes the form of *geomagnetic substorms* (Fig. 2, illustrating this Dungey cycle; Eastwood et al. 2015). Energised plasma returning to the dayside is associated with particle precipitation into the polar regions, where the bright auroral displays are the footprints of the whole interaction (Angelopoulos et al. 2008). The solar wind can also be interrupted by large, fast-moving bursts of plasma called interplanetary coronal mass ejections, or CMEs (Gonzales et al. 1999). When a CME impacts the Earth's magnetosphere it temporarily deforms the Earth's magnetic field, perturbing its direction and strength, and inducing large electrical currents; this is called a *geomagnetic storm* and it is a global phenomenon. CMEs can be associated with prolonged periods of northward or southward IMF, and those with southward IMF are more geoeffective; they represent a severe space weather threat with the greatest capacity to disrupt everyday life throughout the world (e.g. affecting satellite subsystems and astronaut wellbeing in space, telecommunications, electrical infrastructures, and pipelines on the ground).

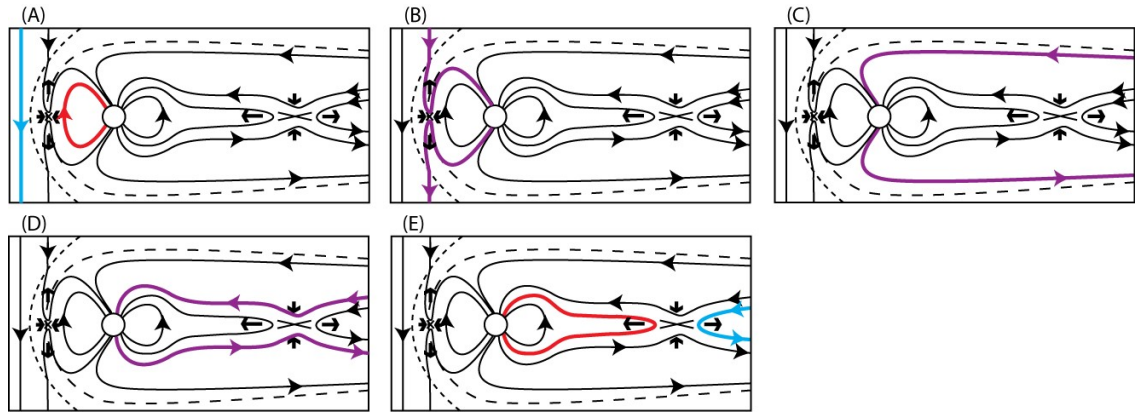

Fig. 2 Cartoon showing the progression of the Dungey cycle. Under southward IMF conditions, dayside reconnection (panel A) opens magnetic flux (panel B) which convects over the poles and is stored as magnetic energy in the magnetotail lobes (panel C). This stored energy accumulates until an explosively release (panel D) returns closed flux to Earth in conjunction with dramatic auroral displays at high latitudes (panel E). (From Eastwood et al. 2015)

## 3 In situ measurements versus global view

As described above, the structure and dynamics of the magnetosphere are mainly controlled by magnetic reconnection. The basic theory of magnetospheric circulation is well known, since the microscale has been explored through many in situ measurements. However, the reality of how this complex interaction takes place on a global scale, and how it evolves with time, is still not fully understood. Moreover, the global models that have been developed to simulate such interaction still require validation.

The interaction between the solar wind and the Earth's magnetosphere, and the geospace dynamics that result, comprise a fundamental aspect of heliophysics. Understanding how this vast system works requires knowledge of energy and mass transport, and a comprehension of coupling between regions and between plasma and neutral populations. Missions providing in situ observations enable us to explore the details of the fundamental microscale processes that drive transport and coupling. In situ instruments on a fleet of solar and solar wind observatories now provide unprecedented observations of these external drivers. Our experimental knowledge of the Earth's magnetosphere has developed thanks to a sequence of increasingly capable satellite missions, such as the four satellite ESA Cluster mission (Escoubet et al. 2001), the Chinese-European Double Star project (Liu et al. 2005), the US Time History of Events and Macroscale Interactions during Substorms (THEMIS, Angelopoulos 2008), the Van Allen Probes (Mauk et al. 2013), the Magnetospheric Multi-Scale (MMS, Burch et al. 2015), and the Japanese Geotail (Mukai et al. 1994) and Arase-ERG (Miyoshi et al. 2018) missions. They have explored the magnetosphere in-situ, making highly precise local measurements of the various plasma processes that control the behaviour of the magnetosphere.

However, we are still unable to quantify the global effects of those drivers, including the conditions that prevail throughout geospace. This information is the key missing link for developing a complete understanding of how the Sun gives rise to and controls Earth's plasma environment. A significant breakthrough in magnetospheric physics came in the form of global auroral imaging, most notably provided by the NASA IMAGE (2000-2005, Burch 2000) and Polar (1996-2008, https://pwg.gsfc.nasa.gov/polar/) missions, which offered a global context for multipoint in-situ spacecraft measurements, and in a more limited fashion from Energetic Neutral Atoms measurements (Mitchell et al. 2001). However, even with global auroral imaging (unfortunately no longer available), we cannot determine in general where the boundaries of the magnetosphere actually lie, because of the inherent uncertainty in magnetic field line mapping. Nor do we have independent confirmation of the physical processes that occur there. Only in fortuitous circumstances are spacecraft appropriately positioned to confirm that a particular process (most specifically reconnection) is occurring in a particular place, at a particular time, to unambiguously explain the auroral observations.

In this context, a novel approach was required and one has been devised from the application of established astronomical techniques to the study of our own planet.

**4 A novel method to image the magnetosphere**

It is a relatively recent discovery (Cox 1998, Cravens 2000) that soft X-rays are produced in the Earth's magnetosheath and magnetospheric cusps by the process of solar wind charge exchange (SWCX, Fig. 3). SWCX takes place when highly charged heavy ion species in the solar wind interact with neutral atoms or molecules. An electron from the neutral is transferred to the ion, which is initially left in a highly excited state. On relaxation to the ground state one or more photons are emitted, with energy matching the atomic transition that has occurred, usually in the extreme ultraviolet or the soft X-ray band. In particular the 0.5- 2.0 keV range is dominated by strong lines due to charge exchange by $O^{7+}$, $O^{8+}$, $Ne^{9+}$, and $Mg^{11+}$.

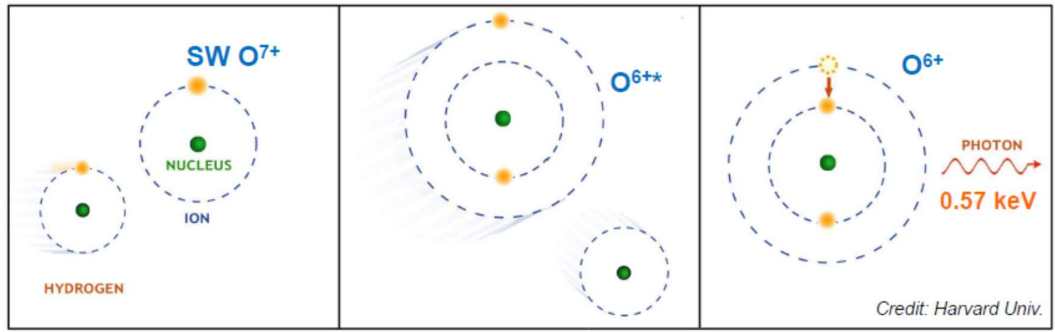

Fig. 3 Schematic illustrating the process of solar wind charge exchange (edited from Chandra X-ray Observatory, Harvard University).

For an in-depth description of the SWCX process and its presence in many astrophysical environments see the review by Sibeck et al. (2018) and also the chapter titled 'Earth's exospheric X-ray emissions' by Jenny Carter in Section VII of this Handbook. There are many sources of SWCX emission in the heliosphere, including comets and the neutral interstellar medium that flows through the solar system. Typically, the brightest source of SWCX is that due to the Earth's exosphere, which is primarily hydrogen, interacting with the shocked, compressed solar wind in the magnetosheath and the magnetospheric cusps. The charge exchange X-ray emissivity is proportional to the density of the highly charged heavy ions and that of the neutrals that undergo the interaction (Cravens 1997), hence the SWCX emission is brightest in the dayside magnetosheath and the cusps. It is interesting to note that, given sufficient instrumental spectral resolution, careful examination of the spectral properties of the X-ray emission not only can provide information of the species and charge states of the solar wind ions, but also of the neutral atoms or molecules involved in the SWCX interaction (e.g. Mullen et al. 2017).

**5 The novel approach with SMILE**

SMILE will investigate the dynamic response of the Earth's magnetosphere to the impact of the solar wind in the global manner so critically needed, and in a way never attempted before. SMILE combines global soft X-ray imaging of the dayside magnetosheath and the cusps with the Soft X-ray Imager (SXI), simultaneous UV imaging of the northern aurora with the UV Imager (UVI) and in situ monitoring of the solar wind and magnetosheath plasma conditions (with the Light Ion Analyser, LIA, and the magnetometer, MAG) from a highly elliptical northern polar orbit that takes it out to an apogee of 20 Earth's radii (Fig. 4). For a detailed description of the SMILE mission and payload readers should refer to the ESA Definition Study Report (Branduardi-Raymont et al. 2018) and the dedicated chapter 'The SMILE mission' in Section III of this Handbook.

SMILE provides the necessary global view, and can answer questions that help distinguishing the modes in which solar-terrestrial interactions take place, the characters of reconnection, what triggers geomagnetic substorms and how CME-driven storms arise. Essentially SMILE will go a long way to answer the question: What drives space weather?

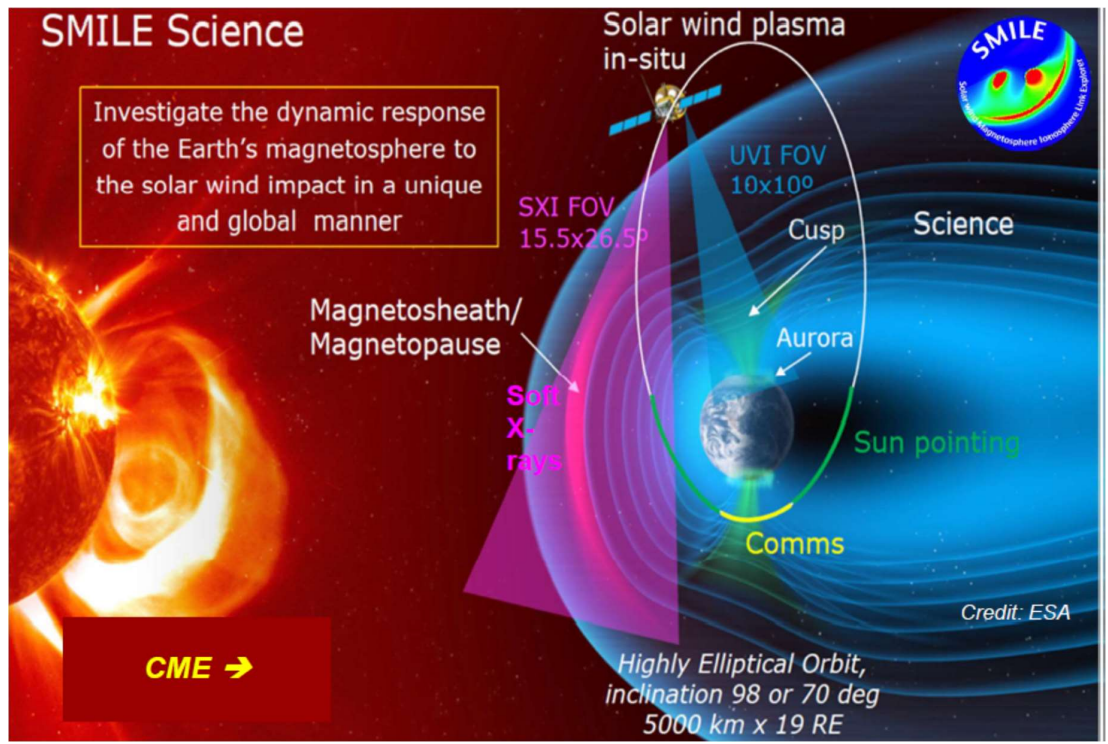

Fig. 4 The global approach of SMILE to investigating solar-terrestrial interactions. Depicted is the case of a coronal mass ejection (CME – shown as the orange arc on the left) travelling towards Earth (see text for details – Credit ESA)

From Fig. 4 it can be seen that SMILE turns X-rays produced in the magnetosheath and the cusps, which vary according to the strength of the solar wind and act as unwanted background for astronomical observations whose line of sight crosses these regions of geospace, into a very valuable diagnostic tool of solar-terrestrial interactions.

**6 SMILE scientific motivations**

We now look in more detail at the science questions which underpin the core motivations of the SMILE mission.

The boundaries seen in soft X-ray (and low energy neutral atom) images correspond to plasma density structures like the bow shock, magnetosheath, magnetopause, and cusps. Thus soft X-ray imagers can be used to track what is called the 'inward erosion' of the dayside magnetopause during the growth phase of geomagnetic substorms and the outward motion of this boundary following substorm onsets. The location of the magnetopause provides information concerning the amount of closed flux within the dayside magnetosphere; the rate of magnetopause erosion or recovery provides information concerning the steadiness of reconnection, while the location of the portion of the magnetopause that moves provides information concerning the component (occurring on the equatorial magnetopause) or antiparallel (at high latitude) nature of reconnection.

Soft X-ray imagers can also be used to track the equatorward motion of the cusps during the substorm growth phase and their poleward motion following onset. Just as in the case of the magnetopause, cusp observations can be used to determine the amount of closed flux within the dayside magnetosphere, the rates of erosion and recovery, the steadiness of reconnection, and the equatorial or high-latitude location of reconnection.

Global auroral images from a high inclination, high altitude spacecraft provide an excellent complement to soft X-ray images. The dimensions of the auroral oval indicate the open magnetic flux within the Earth's magnetotail. Poleward and equatorward motions of the dayside and nightside auroral oval provide crucial information concerning the occurrences and rates of reconnection at the dayside magnetopause and within the Earth's magnetotail.

Finally, measurements of the solar wind plasma and magnetic field input to the magnetosphere by instrumentation located near Earth are essential complements for the above studies, because having such monitors reduces concerns regarding the arrival times of possible solar wind triggers for magnetospheric events and reduces concerns regarding the dimensions of solar wind structures transverse to the Sun-Earth line.

6.1 The character of reconnection

While isolated single or closely-spaced multipoint in situ measurements can be used to identify reconnection events and study the microphysics of reconnection, they cannot be used to distinguish between models in which reconnection is predominantly patchy or global, transient or continuous, triggered by solar wind features or occurring in response to intrinsic current layer instabilities, component and occurring on the equatorial magnetopause or antiparallel and occurring on the high-latitude magnetopause. Nor can isolated measurements be used to determine the global state of the solar wind-magnetosphere interaction, as measured by the rate at which closed magnetic flux is opened or open flux closed. For all of these tasks, and many more, global observations are needed.

In particular, we want to establish the fundamental modes of the dayside solar wind-magnetosphere interactions, and in particular when/where is reconnection steady/transient/bursty, patchy or global. We are interested in finding out when/where any of these conditions dominate. Could they be dependent on the beta parameter, i.e. the ratio of the plasma pressure to the magnetic pressure? It is thought that steady reconnection occurs for low beta (i.e. magnetic field pressure dominated) solar wind and magnetosheath, whereas unsteady reconnection is more likely for high beta solar wind conditions (Phan et al. 2013). However, a simple confirmation of this hypothesis is obscured by the fact that apparently unsteady magnetopause reconnection may simply be directly driven by variations in the solar wind parameters. Systematic measurements, such as those which SMILE offers, will help resolve the issue. As it will also allow us to 'see' and help distinguish cases of component and antiparallel reconnection.

Reconnection may be the cause or consequence of various plasma instabilities proposed to occur within the near-Earth magnetotail. There could also be direct dependence on solar wind parameters. These issues could be settled by correlating in situ data with position, motion and morphology of the magnetopause and cusps using SXI images, and UVI auroral brightenings.

Moreover, what is the role of the magnetospheric cusps in solar wind-magnetosphere coupling? The peculiar magnetic topology of the cusps means that they also play a pivotal role in magnetospheric dynamics: they are the sole locations where solar wind has direct access to low altitudes (e.g. Cargill et al. 2005). They are essentially the boundary that separates magnetic field lines that close in the dayside hemisphere from those that extend far down the magnetotail. During subsolar reconnection solar wind energy, mass and momentum are transferred through the cusps into the magnetosphere.

The latitudinal location of the cusp depends on the level of interconnection of the Earth's dipole with the IMF (Newell et al., 1989). When the solar wind magnetic field points southwards, magnetic reconnection at the magnetopause opens closed dayside magnetic field lines, causing the region of open field lines in the polar cap to expand to lower latitudes. The latitudinal position of the cusp is also an indicator of how much magnetic flux is being removed from the dayside to fuel substorm behaviour on the nightside. When the IMF turns northward, the cusps move poleward. Fig. 5 shows spectrograms gathered by three of the four Cluster spacecraft, that happened to cross the cusps in quick succession during an episode of IMF turning from south to northward (Escoubet et al. 2008). We see that proton energy decreases towards the pole for IMF southward (top two panels), and viceversa for IMF northward (bottom panel). Also, the cusp expands poleward after the IMF turns North. This is a single measurement, occurring by a fortunate coincidence of spacecraft locations. Systematic observations of how the cusp responds to IMF turnings, such as SMILE will start to provide, are needed to follow and interpret these developments more carefully.

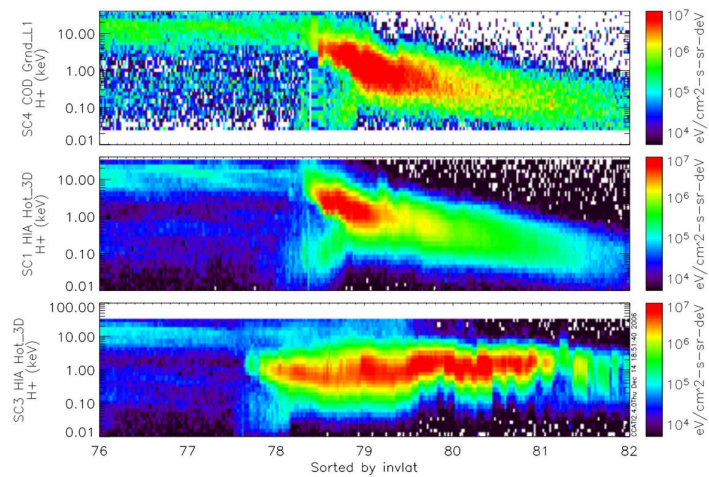

Fig. 5 Cluster 4 and 1 (top two panels respectively) crossed the cusp during southward IMF, 8 min before Cluster 3 (bottom panel) crossed it, after the IMF had turned northward. Proton energy spectra are plotted along the vertical axis in keV, versus geographic latitude in degrees. The colour scale shows proton energy flux (from Escoubet et al. 2008).

Reconnection is thought to cause the shape of the magnetopause to become blunter. By contrast, variations in the solar wind dynamic pressure should cause self-similar changes in magnetospheric dimensions. Thus, by measuring the curvature, size and standoff distance of the magnetopause, and the location (latitudinal position), size and shape of the cusps, it is possible to distinguish the differing effects of pressure changes and magnetic reconnection on the global magnetospheric system. This would distinguish on a global level the nature of the solar wind-magnetosphere interaction, the dominant driving mechanisms and modes of interaction.

6.2 The geomagnetic substorm cycle

We know that southward IMF is required to increase the energy density of the magnetotail lobes, and the more prolonged the interval of southward IMF, the more energy is stored, but the precise nature of the energy loading and the role it plays in the subsequent onset of geomagnetic activity is very controversial. For example, one very fundamental question is whether each substorm requires its own interval of loading (growth phase), or whether multiple substorms can occur in response to a single growth phase. The polar cap is an area of magnetic field lines that are open to the solar wind and is readily identified by the auroral oval which bounds it (Fig. 6). Auroral oval observations provide information about the ionospheric footpoints of magnetopause processes. The trigger that leads to substorm onset remains controversial. Is the substorm triggered by changes in IMF orientation (related to a change in shape of the magnetopause due to reconfiguration of magnetospheric
currents associated with dayside reconnection) (e.g. Hsu and McPherron 2002; Lyons et al. 1997; Morley and Freeman 2007; Wild et al. 2009)? Or do solar wind dynamic pressure changes play the key role (by compressing the magnetotail) (e.g. Boudouridis et al. 2003; Hubert et al. 2006, 2009; Milan et al. 2004)? How large and rapid these driving changes must be is unknown. Another viewpoint is that the external solar wind condition provides only the general configuration of the magnetosphere for substorm expansion onset. When and where it occurs depends on the ionospheric conditions, as well as internal local magnetospheric parameters. Furthermore, the role of the prior history of the magnetosphere in conditioning the response is not well understood and there are reports of substorms with no obvious external drivers (Huang 2002). Are seasonal effects also possibly playing a role?

Thus, despite a plethora of in situ observations, fundamental questions remain unanswered. If the onset of a substorm is due to external driving, what is the nature of the driving mechanism, and how does this depend on the precise configuration of the magnetosphere? With its combined payload made up of the SXI imaging the dayside magnetosheath SWCX X-rays and UVI continuously monitoring the northern auroral oval in the ultraviolet, indeed SMILE explores the link between magnetosphere and ionosphere, and can closely follow the development of geomagnetic substorms.

The auroral oval responds to changes in magnetospheric conditions in a clear fashion: Fig. 6 (from Milan 2009) shows auroral UV images (top panels a and b) captured by the IMAGE spacecraft illustrating the large change in radius (see also panel c) in the auroral oval at the height of a geomagnetic storm and during a following period of magnetospheric quiescence, as quantified by the Sym-H index (a measure of ring current intensity, indicating storms when very negative, panel d) and the rate of solar wind-magnetosphere coupling $\Phi_D$ (derived from upstream measurements of the solar wind speed and the strength and orientation of its embedded magnetic field, panel e). An estimate of the size of the polar cap can be derived from the radius of the auroral oval. It is worth noting that the data gaps in panel c, due to the 14 hour orbital period of IMAGE, hinder the study of storm details and solar wind-magnetosphere coupling. Short time-scale variations in polar cap size correspond to substorms, but large discontinuities exist over some data gaps indicating that the storm behaviour is only partially captured. This is an area where the continuous (~ 40 hours) auroral monitoring by SMILE for the first time will make possible unbroken determination of the rates of magnetic reconnection and of the factors that influence these.

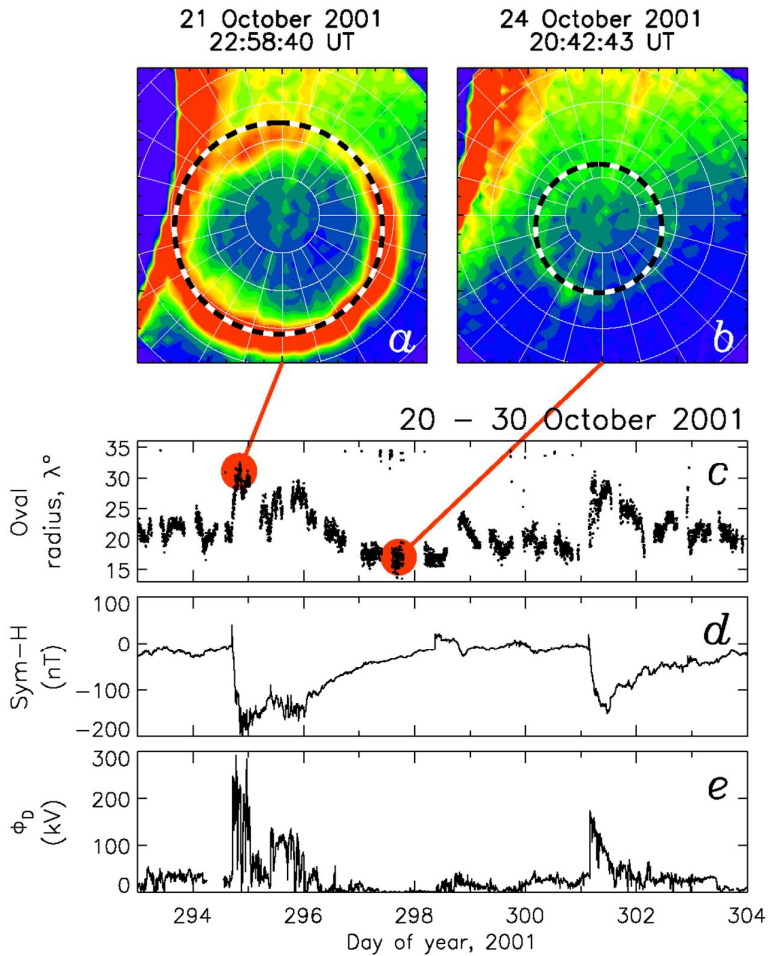

Fig. 6 Changes in the radius of the auroral oval as observed in the UV (panels a, b and c) during and after a geomagnetic storm, as identified by the activity index Sym-H (panel d) and by the rate of solar wind-magnetosphere coupling (panel e). From Milan 2009.

The impact of solar wind variability on the Earth's aurora is not limited to morphological changes of the oval emission distributions. Other modes of magnetospheric behaviour have been observed, such as saw-tooth events (strong magnetic disturbances whose intensity periodically increases and decreases, possibly related to injections of energetic particles at geosynchronous orbit, Walach et al. 2017) and auroral beads (interpreted as caused by instabilities in the magnetospheric configuration preceding substorm onset, Sorathia et al. 2020).

During saw-tooth events, which are oscillations of energetic particle fluxes at geosynchronous orbit recurring with a period of about 2-4 hours (e.g. Henderson et al. 2006), the auroral oval expands and contracts with a period of a few hours. It is not clear if this is due to an intrinsic instability/mode of dynamic behaviour or if it corresponds to a series of repeating substorms. These may simply reflect the same internal physics being driven differently by the solar wind, or they may represent fundamentally different types of behaviour.

A recent development is the recognition of low-intensity auroral features called auroral beads that develop in pre-breakup auroral arcs and eventually produce the initial brightening and substorm expansion onset (Henderson 1994). Auroral beads have specific wavelengths and corresponding exponential growth in the auroral intensity that are different from case to case, apparently dependent on the state of the magnetosphere just prior to substorm expansion onset. The characteristics of auroral beads revealed recently impose another set of rather severe observational constraints that discriminate among several potential substorm onset processes under consideration. Two potential plasma instabilities that may account for these characteristics are the ballooning instability (Sorathia et al. 2020) and the cross-field current instability (Lui 2016). More recently, however, Kalmoni et al. (2018) showed that a third mechanism, kinetic Alfvén waves, could explain the temporal and spatial scales of auroral beads. Coordinated global imaging from SMILE and ground-based auroral observations around substorm expansion onset would be ideal to test these proposed plasma instabilities further. A working group on Ground-Based and Additional Science (GBAS) has indeed been created to coordinate SMILE observations with ground-based measurements (see the chapter 'The SMILE mission' in Section III).

6.3 <u>CME-driven geomagnetic storms</u>

While intervals of southward IMF occur naturally in the solar wind, and so substorms occur on a daily basis (Borovsky et al. 1993), strong solar wind driving causing geomagnetic storms tends to occur in response to coherent solar wind structures, particularly Coronal Mass Ejections (CMEs) (Gonzalez et al. 1999), as depicted in Fig. 4.

CMEs are transient eruptions of material from the Sun's corona into space (Forbes 2000). They often propagate at super-magnetosonic speeds relative to the ambient solar wind, and play a particularly important role in the dynamics of the Earth's magnetic field being associated with long intervals of southward IMF (e.g. Gonzales et al. 1999) which leads to enhanced magnetic reconnection. In general, the largest geomagnetic disturbances are caused by CMEs, with the level of activity being directly related to the flow speed, the field strength and the southward component of the magnetic field (Richardson et al. 2001).

Understanding the global CME-magnetosphere interaction is crucial to understanding precisely how the structure of the CME is responsible for the different phases of geomagnetic storms. On a practical level, storms driven by CMEs have potentially severe space weather consequences and represent a significant threat to infrastructure resilience worldwide. Very basic questions still remain. Is the duration and magnitude of solar wind driving the sole arbiter of whether a storm will occur? What is the relationship between storms and substorms? Are storms always a separate phenomenon, or can they be considered as being composed of multiple substorms?

The question of storms duration is growing in importance, driven by the needs of the end-user in the space weather context (i.e. confidence in issuing 'all clear'). Does a storm end because it has exhausted the reservoir of stored magnetic energy in the magnetotail? Or does a storm stop because the solar wind driving conditions have changed? If both possibilities are observed to occur, which is the more important? And once the solar wind driving is removed, how rapidly does the magnetosphere recover? Is it more likely that the solar wind conditions will change, or is the stored magnetotail lobe energy depleted so rapidly that the changing solar wind plays only a minor role?

The combination of imaging by SMILE SXI and UVI together with the in situ measurements of plasma and magnetic field conditions by LIA and MAG, in the solar wind and the magnetosheath, on a long elliptical orbit, will provide novel global datasets with which to tackle the questions posed above. Such self-sufficient capabilities will be complemented by relating and combining the SMILE data with those from other magnetospheric space missions flying at the same time, and with ground-based measurements, as already mentioned (see also the chapter 'The SMILE mission' in Section III): this is currently being planned and worked on, while new ways of data fusion are also being developed in order to optimise the scientific exploitation of the data that SMILE will return.

## 7 Modelling in preparation for SMILE

Simulation and modelling of the data expected from SMILE, and in particular from the most novel instrument in the payload, the SXI, are advancing at a fast pace. This activity was initiated originally in order to establish the scientific and instrument requirements for the SXI, and has now turned to developing, comparing and optimising techniques of data analysis and parameter estimation that will be required during the operational phases of the mission to extract the most accurate, best science. A most relevant example is the ongoing work in searching for and testing the optimal technique to extract the magnetopause location and shape under differing solar wind conditions.

The kind of outputs produced by this work are illustrated by Fig.s 7a and 7b for the geomagnetic storm that took place on 17 March 2015. Magneto-Hydro-Dynamic (MHD) simulations are carried out to derive the 3-D geomagnetic parameters expected to correspond to given input solar wind conditions (e.g. the quantities plotted in the graphs on the left of the figures). Several MHD codes, such as the 3-D PPMLR (extended Lagrangian version of the piecewise parabolic method, Hu et al. 2007), the BATS-R-US and LFM (both available at the Community Coordinated Modeling Center, or CCMC, website https://ccmc.gsfc.nasa.gov/), are employed and their outputs compared. The 3-D parameters are used in the calculation of the SWCX X-ray emissivity in the magnetosheath and cusps regions, and this is then integrated along SMILE's line of sight along its orbit. The result are 2-D maps of emissivity as shown on the right hand side of Fig.s 7a and 7b. The arc-like region is the magnetosheath which, together with the two magnetospheric cusps, can be seen brightening as the solar wind density and speed increase during the storm.

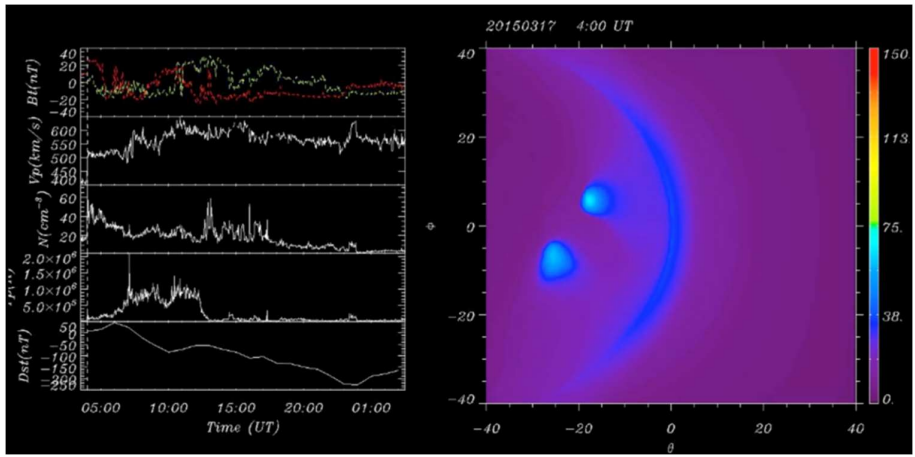

Fig. 7a Simulations of expected SWCX X-ray emissivity (right image) before the onset of the 17 March 2015 storm: 04:00 UT, N = 15 cm$^{-3}$, V = 410 km/s (plotted in the left graph)

Credit: T. Sun, CAS/NSSC

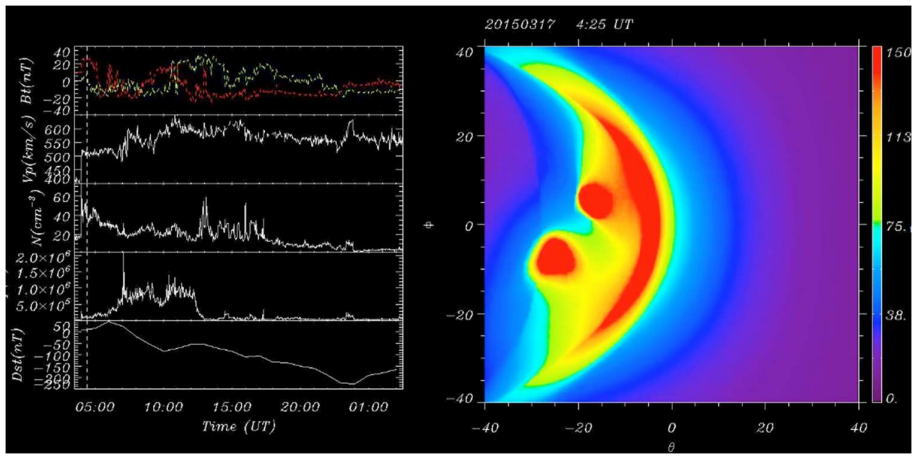

Fig. 7b Simulations of expected SWCX X-ray emissivity (right image) during the 17 March 2015 storm: 04:25 UT, N = 50 cm$^{-3}$, V = 510 km/s (plotted in the left graph)

Credit: T. Sun, CAS/NSSC

The X-ray emissivity maps can then be convolved with the SXI instrument response using the SXI simulator code to produce expected count images such as those the imager will return in flight. An example of the output of this whole process is shown in Fig. 8 at the peak of an event that occurred on 12 September 2014, involving IMF turning from North to South, with high values of solar wind density, speed and magnetic field: on the left is the SWCX X-ray emissivity map and on the right is the expected SXI count image for an integration time of 5 min and after background subtraction.

A number of boundary tracing algorithms have been developed and compared in view of their application in the analysis of SXI images returned during SMILE operational phase. These include the tangential direction approach (TDA) of Collier and Connor (2018) and the tangent fitting approach (TFA) of Sun et al. (2020). On the basis of the limb brightening effect Collier and Connor assert that the soft X-ray peak direction is tangent to the magnetopause and derive an analytical formula of the magnetopause location as a function of spacecraft location and soft X-ray peak angles. Sun et al. construct a series of 2-D X-ray images based on parametric descriptions of the locations of boundaries and parametric expressions for the X-ray brightness distribution between the boundaries, and then find the best match with the 'real' X-ray images by a fitting process of the tangent directions to the positions of maximum emissivity ('real' in the sense that before the launch of SMILE images derived from MHD simulation are used as the 'real' images, and after launch the 'real' X-ray images would be those returned by SXI). An alternative is the computed tomography approach (CTA, Jorgensen et al. 2020) which applies the traditional computerised tomography approach to the X-ray images of the magnetopause. A set of images are collected and analysed and the 3-D X-ray emissivity is reconstructed, with the 3-D magnetopause surface subsequently derived from the emissivity maps. A more direct extraction of the magnetopause location can be obtained by identifying and fitting the magnetopause shape to the locus of maximum X-ray emissivity or that of the maximum emissivity gradient (Samsonov, private comm.). Comparison of the results from the different approaches will ultimately provide an estimate of the uncertainty of the derived location and shape of the magnetopause under changing solar wind conditions.

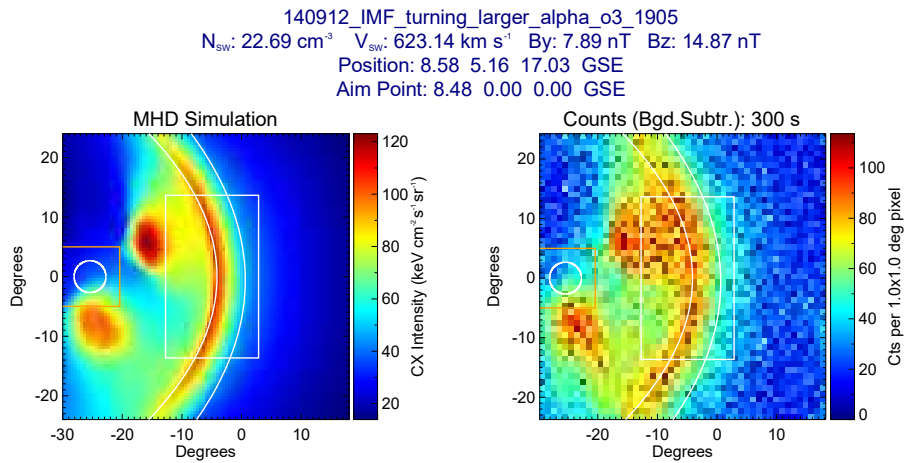

Fig. 8 Left: SWCX X-ray emissivity map from an MHD simulation using the solar wind parameters shown at the top. Right: Background subtracted SXI count image for an integration time of 5 min. The white large rectangular box and the smaller square represent the SXI and UVI fields of view, respectively. The bright arc-like feature is due to the enhanced X-ray emission along the magnetopause, indicated by the inner white curve; the outer white curve shows the position of the bow shock; the magnetosheath lies in between the two. The two bright spots are the magnetospheric cusps.

Credit: S. Sembay, Leicester University

## 8 SMILE impact and conclusions

SMILE is a novel space mission that will revolutionise magnetospheric physics by providing simultaneous images and movies of the Earth's magnetopause, cusps, and auroral oval for up to 40 hours per orbit continuously, using state-of-the-art detection techniques.

Charge exchange is now recognised as a ubiquitous mechanism that produces X-rays throughout the Universe, and the approach adopted with SMILE SXI allows established astronomical techniques to be applied to our own planet for the first time. SMILE SXI will demonstrate how SWCX soft X-rays from the Earth's magnetosheath and magnetospheric cusps, which constitute an unwanted variable soft X-ray background for astrophysical observations of the Universe outside the solar system, can be turned into an important diagnostic tool of solar-terrestrial relationships. An initially challenging and now recognised as rewarding character of SMILE is the coming together of astrophysical, planetary and space plasma communities of researchers: all participants have had to learn lessons in each other's science, in its added value in coming together, in the terminology, the way instruments work, the formats of the data that they return.

The cooperation with China is another interesting aspect of the SMILE mission, this being the first time that ESA and CAS collaborate on a mission from start to finish, from jointly announcing the opportunity of proposing for a space mission, its selection, design and development, to operations and science exploitation. The teams of scientists and engineers collaborating on SMILE have been learning how different cultures may approach issues differently, and how much they have in common and can support each other at the scientific, building and operating levels.

Finally, SMILE has an important contribution to make to public engagement, and much work is already expended in this direction: being a very visual mission, with two imagers onboard, SMILE has strong potential to engage, especially by making visible the so far invisible Earth's magnetic field, together with the associated UV views of the auroral oval. SMILE has the potential to revolutionise the general understanding of this area of science by providing X-ray images of the magnetospheric bubble shielding our Earth from inclement solar wind conditions. The database of SMILE observations will be an unparalleled resource providing a global view and a direct measure of the response of the Earth's magnetospheric system, from its outer dayside boundaries to deep into the ionosphere, under the influx of the solar wind, and especially of changes in the wind conditions. As such it will constitute a golden reference data bank for validation of solar-terrestrial interaction models and for understanding space weather effects, with the ultimate goal of learning how to mitigate them.

*This Chapter will appear in the Section "Solar System" (Section Editor: G. Branduardi-Raymont) of the "Handbook of X-ray and Gamma-ray Astrophysics" by Springer (Editors in chief: C. Bambi and A. Santangelo).*